\documentclass[twocolumn,aps,prl,showpacs,floatfix]{revtex4}
\usepackage{graphicx}
\usepackage{times}
\usepackage{amsmath}
\usepackage{amsfonts}
\usepackage{amssymb}
\usepackage{amsthm}
\usepackage{epsf}
\usepackage{bm}
\usepackage{times}

\usepackage{dcolumn}

\newcolumntype{.}{D{x}{}{-1}}

\newcommand{\la}{\langle}
\newcommand{\ra}{\rangle}

\begin{document}
%%%%%%%%%%%%%%%%%%%%%%%%%%%%%%%%%%%%%%%%%%%%%%%%%%%%%%%%%%%%%%%%%%%%%%%%

\title{
%QED shift calculations in relativistic many-electron atoms and ions
QED shifts in multivalent heavy ions}

\author{I. I. Tupitsyn$^{1,2}$}
\author{M. G. Kozlov$^{2,3}$}
\author{M. S. Safronova$^{4,5}$}
\author{V. M. Shabaev$^{1}$}
\author{V. A. Dzuba$^{6}$}

\affiliation{$^{1}$Department of Physics, St. Petersburg State University,
Ulianovskaya 1, Petrodvorets, St.Petersburg, 198504, Russia}

\affiliation{$^2$Petersburg Nuclear Physics Institute, Gatchina
188300, Russia}

\affiliation{$^3$St.~Petersburg Electrotechnical University
``LETI'', Prof. Popov Str. 5, St.~Petersburg, 197376, Russia}

\affiliation{$^{4}$Department of Physics and Astronomy, University
of Delaware, Newark, Delaware 19716, USA}

\affiliation{$^{5}$Joint Quantum Institute, National Institute of
Standards and Technology and the University of Maryland,
Gaithersburg, Maryland 20742, USA}

\affiliation{$^6$School of Physics, University of New South Wales, Sydney 2052, Australia}

\begin{abstract}
The quantum electrodynamics (QED) corrections are directly incorporated 
into the most accurate treatment of the correlation corrections for ions 
with complex electronic structure of interest to metrology and tests of 
fundamental physics. We compared the performance of four different QED 
potentials for various systems to access the accuracy of QED calculations 
and to make prediction of highly charged ion properties urgently needed 
for planning future experiments.  We find that all four potentials give 
consistent and reliable results for ions of interest. For the strongly 
bound electrons the nonlocal potentials are more accurate than the local potential.
%We incorporated quantum electrodynamics (QED) corrections into the
%broadly-applicable high-precision relativistic method that combines
%configuration interaction (CI) and linearized coupled-cluster
%approaches. With the addition of the QED, this CI+all-order method
%allows one to accurately predict properties of heavy ions of
%particular interest to the design of precision atomic clocks and
%tests of fundamental physics. To evaluate the accuracy of the QED
%contributions and test various QED models, we incorporated four
%different one-electron QED potentials. We demonstrated that all of
%them give consistent and reliable results. For the strongly bound
%electrons (i.e. inner electrons of heavy atoms, or valence electrons
%in highly-charged ions), the nonlocal potentials are more accurate,
%than the local one [Flambaum and Ginges, Phys.~Rev. A {\bf 72},
%052115 (2005)]. Results are presented for cases of particular
%experimental interest.
\end{abstract}
\pacs{31.30.J-, 12.20.Ds}
\maketitle
%%%%%%%%%%%%%%%%%%%%%%%%%%%%%%%%%%%%%%%%%%%%%%%%%%%%%%%%%%%%%%%%%%%%
%\section{Introduction}
%%%%%%%%%%%%%%%%%%%%%%%%%%%%%%%%%%%%%%%%%%%%%%%%%%%%%%%%%%%%%%%%%%%%%

Optical transitions in heavy many-electron highly charged ions (HCI)
have been recently proposed for the development of ultra-precision
atomic clocks and tests of fundamental physics
\cite{BDF10,BerDzuFla11b,lli,SafDzuFla14,DzuSafSaf15}. From the
experimental standpoint, locating these ultra-narrow optical
transitions is particularly difficult. For most of these ions, with
the degrees of ionization ranging from ${8^+}$ to ${18^+}$, no
experimental data exist and identification of their complicated
atomic spectra is a very difficult task \cite{WLBO15} unless
accurate theoretical predictions are available. Therefore, it is
crucial to develop methodologies for reliable prediction of their
properties for rapid experimental progress toward the new
applications.

In 2015, sympathetic cooling of Ar$^{13+}$ with laser cooled Be$^+$
ions have been demonstrated \cite{SchVerSch15}, elevating HCIs to
the realm of applications previously limited to singly-charged ions
currently used for atomic clocks \cite{Al}, quantum information
\cite{quant}, and other applications requiring laser cooling and
trapping. Accurate prediction of wavelength of optical transitions
suitable for clock development is a  difficult task due to very
large cancelations of the energies of upper and lower states. In
these ions, high-order correlation, Breit, and radiative quantum
electrodynamic (QED) corrections are all important, with cancelation
of these contributions making accurate computations even more
difficult \cite{SafDzuFla14}. As a result, it has become urgent to
accurately take into account QED corrections in calculations of the
electronic structure of such many-electron ions.

Non-empirical calculations of radiative corrections using the QED
perturbation theory for many-electron systems are extremely
complicated and time-consuming. To date, all-order high-accuracy
calculations can be performed only for highly-charged few-electron
ions (see, e.g.,
\cite{johnson_85,mohr_92a,mohr_92b,Perss_96,lab_99,yer99,
art05,yer06,art07,koz10,sap_11,vol12,mal15,yer15} and references
therein), or using the same perturbative methods for many-electron
systems, but with an effective screening potential
\cite{sap_02,sap_03,sha05a,sha05b,che_06,sap_15}. This potential can
be constructed using Dirac-Hartree and Dirac-Fock-Slater (DFS)
methods, or density functional theory (DFT) in the local density
approximation (LDA). {\it Ab initio} QED methods are too complicated
to be directly incorporated into the Dirac-Coulomb-Breit (DCB)
many-electron calculations. For this reason, numerous attempts have
been undertaken to propose simple methods for incorporating QED
corrections into the many-configuration Dirac-Fock (MCDF),
configuration interaction Dirac-Fock,
%(CI-DF)
and relativistic many-body perturbation theory (MBPT) codes (see,
e.g.,
\cite{ind_90,pyy_03,dra_03,fla_05,thi_10,pyy_12,Tup_13,low13,rob_13,
Shab_13,Ginges_16a,Ginges_16b} and references therein).

In this work, we combined the most accurate treatment of the
correlation corrections for multivalent atoms \cite{SafKozJoh09}
with four different QED potentials, which allows us for the first
time to accurately calculate and systematically study QED corrections in heavy ions with
complex electronic structure. To check the accuracy of all these
potentials we also calculated self-energy (SE) corrections to the
one-electron energies of the valence states of the neutral alkali
metals and to the transition energies in Cu-like ions and compared
our results with the {\it ab initio} calculations.

We selected three representative HCIs with different electronic
configurations as the test cases for the QED contributions to the
DCB Hamiltonian. All of these ions were included  in the studies of
the applications of HCIs to the development of clocks and tests of
the variation of the fundamental constants
\cite{SafDzuFla14,SafDzuFla14a,SafDzuFla14b,DzuSafSaf15}. Ba$^{8+}$
was selected owing to the availability of the experimental values
for comparison, Eu$^{14+}$ was chosen as the test case with the
$f^3$ configuration, and Cf$^{15+}$ has the largest sensitivity to
the alpha-variation in a system which satisfies all the requirement
for the development of accurate optical atomic clocks
\cite{DzuSafSaf15}.

We
%carry out the calculations using
use a high-precision relativistic hybrid approach that combines
configuration interaction and a linearized variant of the
single-double coupled-cluster method, generally referred to as
CI+all-order approach \cite{SafKozJoh09}. This method allows to
include dominant correlation correction to all orders of
perturbation theory. Breit  corrections were included into the
calculations. To separate the QED corrections, the CI+all-order
computations were carried out with and without the QED corrections
and difference was taken to be the QED contribution.

The main goals of this study were to answer the following questions
for the type of ions that are of interest to the applications
mentioned above, i.e.\ many-electron ions with a few valence
electrons:
\\
%$\bullet$
(i) How important is QED correction for accurate
prediction of the energy levels of these ions for future
experimental exploration?
\\
%$\bullet$
(ii) How much the QED correction depends on the version of
the model potential being used?
\\
%$\bullet$
(iii) Is it important to include the QED correction when
constructing the basis set orbitals?
\\
%$\bullet$
(iv) Does QED contribution in such many-electron system depend on
the accuracy of the inclusion of the correlation corrections, i.e.
will the QED corrections calculated in the CI+MBPT and CI+all-order
approximations differ?

\vspace{1mm}
%
%\begin{itemize}
%\item
%How important is QED correction for accurate prediction of the
%energy levels of these ions for future experimental exploration?
%\item
%How much the QED correction depend on the version of the model
%potential being used?
%\item
%Is it important to include the QED correction when constructing the
%basis set orbitals?
%\item
%Does QED contribution in such many-electron system depends on the
%accuracy of the inclusion of the correlation corrections, i.e. will
%the QED corrections calculated in the  CI+MBPT and CI+all-order
%approximations differ?
%\end{itemize}

%%%%%%%%%%%%%%%%%%%%%%%%%%%%%%%%%%%%%%%%%%%%%%%%%%%%%%%%%%%%%%%%%%%%
%\section{One-electron QED potential}
%%%%%%%%%%%%%%%%%%%%%%%%%%%%%%%%%%%%%%%%%%%%%%%%%%%%%%%%%%%%%%%%%%%%%
We present one-electron QED potential as a following sum
%of three parts
%
\begin{equation}
V^{\rm QED} = V^{\rm SE} + V_{\rm Uehl} +  V_{\rm WK} \,,
\end{equation}
where $V^{\rm SE}$ is the self-energy operator, $V_{\rm Uehl}$ and
$V_{\rm WK}$ are the Uehling and Wichmann-Kroll parts of the vacuum
polarization respectively.
%$V_{\rm Uehl}$ is the Uehling part of the vacuum polarization and
%$V_{\rm WK}$ is the Wichmann-Kroll part of vacuum polarization.
Both $V_{\rm Uehl}$ and $V_{\rm WK}$ are local potentials, so their
treatment is rather straightforward and is the same in all four
versions of the calculations, which differ by the treatment of the
SE potential. The Uehling potential can be evaluated by a direct
numerical integration of the well-known formula \cite{Uehl_35}, or,
more easily, by using the approximate formulas from Ref.\
\cite{full_76}. A direct numerical evaluation of the Wichmann-Kroll
potential $V_{\rm WK}$ is rather complicated. For the purpose of the
present work, it is sufficient to use the approximate formulas for
the point-like  nucleus from Ref.~\cite{fai_91}.

%%%%%%%%%%%%%%%%%%%%%%%%%%%%%%%%%%%%%%%%%%%%%%%%%%%%%%%%%%%%%%%%%%%%
\paragraph{Method M1. Model self-energy potential.}
%%%%%%%%%%%%%%%%%%%%%%%%%%%%%%%%%%%%%%%%%%%%%%%%%%%%%%%%%%%%%%%%%%%%%
Following \cite{Shab_13, Shab_15} we approximate the one-electron SE
operator as the sum of local $V^{\rm SE}_{\rm loc}$ and nonlocal
$V_{\rm nl}$ potentials
\begin{eqnarray}
V^{\rm SE} = V^{\rm SE}_{\rm loc} + V_{\rm nl} \,,
\label{se1}
\end{eqnarray}
where nonlocal potential is given in a separable form
\begin{eqnarray}
V_{\rm nl} =
\sum_{i,k=1}^{n} |\phi_i\ra B_{ik}\la \phi_k|\,,
\label{nl1}
\end{eqnarray}
Here $\phi_i$ are so-called projector functions. The choice of these
functions depends on the method of construction of the nonlocal
potential $V_{\rm nl}$ and is described in details in
\cite{Shab_13}.
%for the present case.
%The matrix elements $B_{ik}$ satisfy the following property:
The constants $B_{ik}$ are chosen so that the matrix elements of the
model operator $V^{\rm SE}_{ik}$ calculated with hydrogen like wave
functions $\psi_i$ have to
%exactly coincide with
be equal to matrix elements $Q_{ik}$ of the symmetrized exact
one-loop energy-dependent SE operator $\Sigma(\varepsilon)$
\cite{shab_93}:
\begin{eqnarray}
\la \psi_i | V^{\rm SE} | \psi_k \ra = Q_{ik} =\tfrac{1}{2} \, \left
[ \Sigma(\varepsilon_i)+ \Sigma(\varepsilon_k) \right ]\,.
\end{eqnarray}
%%
%\begin{eqnarray}
%Q_{ij} =\tfrac{1}{2} \, \left [ \Sigma(\varepsilon_i)+
%\Sigma(\varepsilon_k) \right ]
%\end{eqnarray}
%%
%The diagonal and non-diagonal matrix elements $Q_{if}$ of the
%symmetrized one-loop SE operator are presented in \cite{Shab_13}.
%The matrix elements $B_{ik}$ are determined from
%%
%\begin{eqnarray}
%\la \psi_i | V^{\rm SE} | \psi_k \ra = Q_{ik}
%\end{eqnarray}
%%
Introducing two matrices
$\Delta Q_{ik}=Q_{ik}- \la \psi_i|V^{\rm SE}_{\rm loc}| \psi_k\ra$
and $D_{ik} = \la \phi_i|\psi_k\ra $,
we find that
\begin{eqnarray}
B_{ik} = \sum_{j,l=1}^{n} (D^{-1})_{ji}
\la \psi_j| \Delta Q_{jl} | \psi_l\ra (D^{-1})_{lk} \,.
\label{nl2}
\end{eqnarray}
The local part of the SE potential in \cite{Shab_13} was taken in
the simple form
\begin{eqnarray}
 V^{\rm SE}_{{\rm loc},\kappa}(r) =  A_{\kappa} \exp{(-r/\lambdabar_C)}\,,
\label{local}
\end{eqnarray}
where the constant $A_{\kappa}$ is chosen to reproduce
the  SE shift for the lowest energy level at the given $\kappa$
in the corresponding H-like ion, and $\lambdabar_C=\hbar/(mc)$.
The computation code based on this method is presented in
Ref. \cite{Shab_15}.
%%%%%%%%%%%%%%%%%%%%%%%%%%%%%%%%%%%%%%%%%%%%%%%%%%%%%%%%%%%%%%%%%%%%
%\\

%%%%%%%%%%%%%%%%%%%%%%%%%%%%%%%%%%%%%%%%%%%%%%%%%%%%%%%%%%%%%%%%%%%%%%%%%%
\begin{table}[tb]
\begin{center}
\caption{ The SE function $F(\alpha Z)$ for some
neutral alkali metals calculated by methods M1 --- M4. Raws
``Exact'' present \textit{ab initio} results from Ref.\
\cite{sap_02}.}
\label{alkali}
\vspace{-0mm}
%\begin{ruledtabular}

%\end{ruledtabular}
\begin{tabular}{cccccc}
\hline \hline
&&&&&\\[-2mm]
%\hspace{10mm}
%& \hspace{10mm} & DFS & DFS & DFS & DFS & \hspace{2mm} DF \hspace{5mm}\\[-2mm]
%
Atom & Method  & \hspace{0.5mm}  $x_{\alpha}=0$   \hspace{0.5mm}
&   \hspace{0.5mm}  $x_{\alpha}=1/3$ \hspace{0.5mm}
&   \hspace{0.5mm}  $x_{\alpha}=2/3$ \hspace{0.5mm}
&   \hspace{0.5mm}  $x_{\alpha}=1$ \hspace{0.5mm}
   \\[0mm]
\hline
&&&&& \\[-1mm]
Na $3s_{1/2}$ & M1    & 0.1695~ & 0.1675~ & 0.1829~ & 0.2239~  \\[0mm]
              & M2    & 0.1690~ & 0.1671~ & 0.1826~ & 0.2237~  \\[0mm]
              & M3    & 0.1797~ & 0.1763~ & 0.1911~ & 0.2324~  \\[0mm]
              & M4    & 0.1729~ & 0.1699~ & 0.1848~ & 0.2253~  \\[0mm]
              & Exact & 0.1693~ & 0.1674~ & 0.1814~ & 0.2233~  \\[2mm]
K  $4s_{1/2}$ & M1    & 0.0721~ & 0.0723~ & 0.0827~ & 0.1095~  \\[0mm]
              & M2    & 0.0718~ & 0.0721~ & 0.0826~ & 0.1094~  \\[0mm]
              & M3    & 0.0753~ & 0.0752~ & 0.0856~ & 0.1128~  \\[0mm]
              & M4    & 0.0728~ & 0.0728~ & 0.0831~ & 0.1098~  \\[0mm]
              & Exact & 0.0720~ & 0.0721~ & 0.0829~ & 0.1097~  \\[2mm]
Rb $5s_{1/2}$ & M1    & 0.0229~ & 0.0237~ & 0.0284~ & 0.0397~ \\[0mm]
              & M2    & 0.0228~ & 0.0236~ & 0.0284~ & 0.0396~ \\[0mm]
              & M3    & 0.0236~ & 0.0244~ & 0.0292~ & 0.0407~ \\[0mm]
              & M4    & 0.0229~ & 0.0237~ & 0.0284~ & 0.0397~ \\[0mm]
              & Exact & 0.0228~ & 0.0236~ & 0.0283~ & 0.0396~ \\[2mm]
Cs $6s_{1/2}$ & M1    & 0.0127~ & 0.0132~ & 0.0163~ & 0.0236~ \\[0mm]
              & M2    & 0.0126~ & 0.0132~ & 0.0162~ & 0.0236~ \\[0mm]
              & M3    & 0.0130~ & 0.0136~ & 0.0166~ & 0.0241~ \\[0mm]
              & M4    & 0.0127~ & 0.0132~ & 0.0163~ & 0.0236~ \\[0mm]
              & Exact & 0.0126~ & 0.0132~ & 0.0162~ & 0.0235~ \\[2mm]
Fr $7s_{1/2}$ & M1    & 0.0069~ & 0.0076~ & 0.0099~ & 0.0151~ \\[0mm]
              & M2    & 0.0069~ & 0.0076~ & 0.0099~ & 0.0151~ \\[0mm]
              & M3    & 0.0070~ & 0.0077~ & 0.0100~ & 0.0153~ \\[0mm]
              & M4    & 0.0069~ & 0.0076~ & 0.0099~ & 0.0151~ \\[0mm]
              & Exact & 0.0068~ & 0.0075~ & 0.0098~ & 0.0150~ \\[2mm]
\hline
\hline
\vspace{-12mm}
\end{tabular}
\end{center}
\end{table}
%
%%%%%%%%%%%%%%%%%%%%%%%%%%%%%%%%%%%%%%%%%%%%%%%%%%%%%%%%%%%%%%%%%%%%
\paragraph{Method M2. Self-energy nonlocal potential.}
%%%%%%%%%%%%%%%%%%%%%%%%%%%%%%%%%%%%%%%%%%%%%%%%%%%%%%%%%%%%%%%%%%%%%
%
In this approach we use the same equations (\ref{se1}), (\ref{nl1}),
(\ref{nl2}) to construct the SE potential, but use radiative
potential developed in \cite{Berest_82, fla_05} as the local part.
In \cite{fla_05}, the self-energy part of the total radiative
potential is divided into three terms:
\begin{eqnarray}
V^{\rm SE}_{\rm loc}=\Phi_{\rm rad} = \Phi_{\rm mag} +
\Phi_{\rm lf} + \Phi_{\rm hf},
\label{rad1}
\end{eqnarray}
where the potentials $\Phi_{\rm mag}$, $\Phi_{\rm lf}$ and
$\Phi_{\rm hf}$ are referred to as the
magnetic form factor, the low- and high-frequency parts of the
electric form factor, respectively, according to \cite{fla_05} . The expressions for these
potentials are given by Eqs.\ (7, 9, 10) in \cite{fla_05}. Then, we obtain for
the total SE potential
\begin{eqnarray}
V^{\rm SE}=\Phi_{\rm rad}+ \sum_{i,k=1}^{n} | \phi_i\ra \, B_{ik} \,
\la \phi_k|\,. \label{se2}
\end{eqnarray}
The electric form factor contains some fitting parameters to
reproduce the SE corrections for $5s$ and $5p$ states of heavy
H-like ions. However the local radiative potential $\Phi_{\rm rad}$
gives the SE contribution for the $1s$ state with only 10\% accuracy
\cite{fla_05} (see method M3 below). The SE potential (\ref{se2})
which contains the nonlocal part in addition to the local radiative
potential reproduces the low lying SE corrections of the H-like ions
exactly.

%%%%%%%%%%%%%%%%%%%%%%%%%%%%%%%%%%%%%%%%%%%%%%%%%%%%%%%%%%%%%%%%%%%%
\paragraph{Method M3. Local radiation potential.}
%%%%%%%%%%%%%%%%%%%%%%%%%%%%%%%%%%%%%%%%%%%%%%%%%%%%%%%%%%%%%%%%%%%%
Here, we neglect the nonlocal term in \eqref{se2} and use local
radiative potential $V^{\rm SE}=V^{\rm SE}_{\rm loc}=\Phi_{\rm rad}$
from Eq.\ \eqref{rad1} as a full SE one-electron potential
\cite{fla_05}.
%%
%\begin{eqnarray}
%V^{\rm SE}=V^{\rm SE}_{\rm loc}=\Phi_{\rm rad} = \Phi_{\rm mag} +
%\Phi_{\rm lf} + \Phi_{\rm hf}.
%\label{rad2}
%\end{eqnarray}
%%
This radiative potential was widely used in many-electron
calculations, for example, see
\cite{Dinh_08,thi_10,rob_13,Ginges_16b} and references therein. Note
that this local potential was optimized for weakly bound valence
states of heavy neutral atoms and may be less accurate for strongly
bound ionic, or core states.

%%%%%%%%%%%%%%%%%%%%%%%%%%%%%%%%%%%%%%%%%%%%%%%%%%%%%%%%%%%%%%%%%%%%
\paragraph{Method M4. Nonlocal self-energy potential.}
%%%%%%%%%%%%%%%%%%%%%%%%%%%%%%%%%%%%%%%%%%%%%%%%%%%%%%%%%%%%%%%%%%%%%
This approach developed in \cite{Tup_13} is similar to method M2,
but is simpler: it uses only diagonal matrix elements $Q_{ii}$ of
the exact one-loop SE operator $\Sigma(\varepsilon)$ and different
projector functions: %%
\begin{eqnarray}
V^{\rm SE}=V^{\rm SE}_{\rm loc}+ \sum_{i,k=1}^{n} | \phi_i\ra \,
B^{\prime}_{ik} \, \la \phi_k|\,, \label{se3}
\end{eqnarray}
where $V^{\rm SE}_{\rm loc}=\Phi_\mathrm{rad}$ and $\phi_i = V^{\rm
SE}_{\rm loc} \, \psi_i$.
The expectation value of this potential, calculated with the wave
functions $\psi_i$ of H-like ions is equal to the self-energy
corrections $Q_{ii}$:
 $\la \psi_i | V^{\rm SE} | \psi_i \ra = Q_{ii}$.
Coefficients $B^{\prime}_{kj}$ were obtained in \cite{Tup_13}:
\begin{eqnarray}
B^{\prime}_{kj} = \frac{1}{2} \, \left[ \frac{\Delta
Q_{ii}}{D_{ii}}+\frac{\Delta Q_{jj}}{D_{jj}} \right] \,
(D^{-1})_{ij}\,,
\end{eqnarray}
where $\Delta Q_{ii}= Q_{ii}-D_{ii}$ and $D_{ij} =\la \psi_i |
\Phi_{\rm rad} | \psi_j \ra $.

%%%%%%%%%%%%%%%%%%%%%%%%%%%%%%%%%%%%%%%%%%%%%%%%%%%%%%%%%%%%%%%%%%%%
\begin{table}[tb]
\begin{center}
\vspace{1mm} \caption{The SE correction to the $4s-4p$, $4p-4d$,
transition energies in Cu-like ions (eV) for $x_{\alpha} = 2/3$.
Raws ``Exact'' present \textit{ab initio} results from Ref.\
\cite{che_06}.}
\label{Cu-like} \vspace{-0mm}
%\begin{ruledtabular}

%\end{ruledtabular}
\begin{tabular}{ccccccc}
\hline \hline
&&&&&&\\[-2mm]
Ion & Method  &  4s-4p$_\frac{1}{2}$ & 4s-4p$_\frac{3}{2}$ &
4p$_\frac{1}{2}$-4d$_\frac{3}{2}$ &
4p$_\frac{3}{2}$-4d$_\frac{3}{2}$ &
4p$_\frac{3}{2}$-4d$_\frac{5}{2}$  \\[1mm]
\hline
&&&&&& \\[-1mm]
Yb$^{41+}$    & M1    &$ -1.28$~ &$ -1.21$~ &$ -0.11$~ &$ -0.18$ &$ -0.14$ \\[0mm]
              & M2    &$ -1.28$~ &$ -1.20$~ &$ -0.11$~ &$ -0.18$ &$ -0.14$ \\[0mm]
              & M3    &$ -1.28$~ &$ -1.21$~ &$ -0.12$~ &$ -0.19$ &$ -0.15$ \\[0mm]
              & M4    &$ -1.28$~ &$ -1.20$~ &$ -0.11$~ &$ -0.19$ &$ -0.14$ \\[0mm]
              & Exact &$ -1.28$~ &$ -1.21$~ &$ -0.11$~ &$ -0.18$ &$ -0.14$ \\[2mm]
Au$^{50+}$    & M1    &$ -2.17$~ &$ -2.10$~ & 0.28~ &$ -0.35$~ &$ -0.27$ \\[0mm]
              & M2    &$ -2.17$~ &$ -2.09$~ & 0.28~ &$ -0.35$~ &$ -0.27$ \\[0mm]
              & M3    &$ -2.16$~ &$ -2.08$~ & 0.29~ &$ -0.36$~ &$ -0.29$ \\[0mm]
              & M4    &$ -2.17$~ &$ -2.09$~ & 0.28~ &$ -0.36$~ &$ -0.28$ \\[0mm]
              & Exact &$ -2.18$~ &$ -2.10$~ & 0.28~ &$ -0.35$~ &$ -0.28$ \\[2mm]
Bi$^{54+}$    & M1    &$ -2.69$~ &$ -2.64$~ & 0.41~ &$ -0.46$~ &$ -0.36$ \\[0mm]
              & M2    &$ -2.69$~ &$ -2.64$~ & 0.40~ &$ -0.46$~ &$ -0.36$ \\[0mm]
              & M3    &$ -2.67$~ &$ -2.61$~ & 0.42~ &$ -0.48$~ &$ -0.38$ \\[0mm]
              & M4    &$ -2.70$~ &$ -2.63$~ & 0.41~ &$ -0.47$~ &$ -0.36$ \\[0mm]
              & Exact &$ -2.70$~ &$ -2.64$~ & 0.40~ &$ -0.46$~ &$ -0.37$ \\[2mm]
Th$^{61+}$    & M1    &$ -3.84$~ &$ -3.89$~ & 0.75~ &$ -0.71$~ &$ -0.56$ \\[0mm]
              & M2    &$ -3.84$~ &$ -3.88$~ & 0.75~ &$ -0.71$~ &$ -0.56$ \\[0mm]
              & M3    &$ -3.76$~ &$ -3.82$~ & 0.81~ &$ -0.75$~ &$ -0.62$ \\[0mm]
              & M4    &$ -3.84$~ &$ -3.88$~ & 0.75~ &$ -0.72$~ &$ -0.56$ \\[0mm]
              & Exact &$ -3.85$~ &$ -3.89$~ & 0.74~ &$ -0.71$~ &$ -0.57$ \\[2mm]
U$^{63+}$     & M1    &$ -4.22$~ &$ -4.33$~ & 0.90~ &$ -0.79$~ &$ -0.63$ \\[0mm]
              & M2    &$ -4.23$~ &$ -4.32$~ & 0.89~ &$ -0.79$~ &$ -0.63$ \\[0mm]
              & M3    &$ -4.12$~ &$ -4.24$~ & 0.97~ &$ -0.85$~ &$ -0.71$ \\[0mm]
              & M4    &$ -4.23$~ &$ -4.32$~ & 0.89~ &$ -0.80$~ &$ -0.64$ \\[0mm]
              & Exact &$ -4.24$~ &$ -4.33$~ & 0.88~ &$ -0.79$~ &$ -0.65$ \\[2mm]
\hline
\hline
\end{tabular}
\end{center}
\end{table}
%
%%%%%%%%%%%%%%%%%%%%%%%%%%%%%%%%%%%%%%%%%%%%%%%%%%%%%%%%%%%%%%%%%%%%
\begin{table}
\caption{\label{qed1}
%%%%%%%%%%%%%%%%%%%%%%%%%%%%%%%%%%%%%%%%%%%%%%%%%%%%%%%%%%%%%%%%%%%%
Comparison of the QED corrections obtained using methods M1 -- M4 to
the energies of Ba$^{8+}$\!, Eu$^{14+}$\!, and Cf$^{15+}$
calculated in the CI+all-order approach (cm$^{-1}$). Column labelled
M1$^\prime$ gives results of the CI+MBPT calculation.
Column labelled CI-M1 gives results of the calculation  where QED potential was
included only in CI Hamiltonian. First variant of the QED potential (M1) was used in both of these calculations.}
%\caption{\label{qed1}
%%%%%%%%%%%%%%%%%%%%%%%%%%%%%%%%%%%%%%%%%%%%%%%%%%%%%%%%%%%%%%%%%%%%%
%Comparison of the QED corrections to the energies  of Ba$^{8+}$,
%Eu$^{14+}$,  Cf$^{15+}$ obtained using four different QED potentials,
%M1 - M4 in cm$^{-1}$. The results in column labeled ``CI-M1''
%are obtained by including first version of the QED potential only in the CI
%Hamiltonian. The results in the M1$^\prime$ column are obtained using CI+MBPT
%method and the M1 potential, all other results are calculated in
%the CI+all-order approach.}
\begin{ruledtabular}
\begin{tabular}{llcrrrrrr}
\multicolumn{1}{c}{Ion}& \multicolumn{1}{c}{Conf.}&
\multicolumn{1}{c}{Term}&
\multicolumn{1}{r}{CI-M1}&\multicolumn{1}{r}{M1$^\prime$}&
\multicolumn{1}{r}{M1} & \multicolumn{1}{r}{M2} &\multicolumn{1}{r}{M3} &
\multicolumn{1}{r}{M4}\\
\hline
Ba$^{8+}$& $ 5s^2  $&  $^1S_0  $& 974    &972  &  965    &    955    &    987    &    964    \\
       & $ 5p^2  $&  $^3P_0    $& 28     &-30  &  -31    &    -34    &    -24    &    -33    \\
       & $ 5p^2  $&  $^3P_1    $& 56     &5    &  4      &    2      &    13    &    4    \\
       & $ 5p^2  $&  $^3P_2    $& 78     &25   &   27     &    25     &    36    &    27    \\
       & $ 5p^2  $&  $^1D_2    $& 113    &69   &  69     &    69     &    81    &    71    \\
       & $ 5p^2  $&  $^1S_0    $& 98     &52   &  51     &    51     &    62    &    53    \\
       & $ 5s5d  $&  $^3D_1    $& 484    &459  &  455    &    449    &    464    &    453   \\ [0.3pc]
       & $ 5s5p  $&  $^3P_0    $& 503    &471  &  469    &    462    &    483    &    467    \\
       & $ 5s5p  $&  $^1P_1    $& 538    &513  &  508    &    503    &    524    &    508    \\
       & $ 4f5s  $&  $^3F_2    $& 472    &438  &  435    &    430    &    439    &    434    \\
       & $ 4f5s  $&  $^1F_3    $& 462    &424  &  421    &    416    &    425    &    420    \\[0.5pc]
Eu$^{14+}$& $ 4f^2 6s$& $ 3.5    $& 1025&780  &  778    &    766    &    762    &    774    \\
       & $ 4f^2 6s$& $ 4.5    $& 1024  & 779  & 777    &    766    &    761    &    773    \\
       & $ 4f^2 6s$& $ 5.5    $& 1025  & 781  & 779    &    768    &    764    &    775    \\
       & $ 4f^2 6s$& $ 1.5    $& 1025  & 781  & 778    &    767    &    763    &    775   \\ [0.3pc]
       & $ 4f^3   $& $ 4.5    $& 0    &  -426 & -421    &    -420    &    -474    &    -424    \\
       & $ 4f^3   $& $ 5.5    $& 0    &  -425 & -420    &    -419    &    -473    &    -423    \\
       & $ 4f^3   $& $ 6.5    $& 0    &  -424 & -419    &    -418    &    -472    &    -423   \\ [0.5pc]
Cf$^{15+}$  & $ 5f6p^2 $& $ ^2F_{5/2}$& 828 & -265 &  -238    &    -249    &    -178    &    -266    \\
       & $ 5f^26p $& $ ^4I_{9/2}  $& 431    & -781 &  -762    &    -769    &    -815    &    -788    \\
       & $ 5f6p^2 $& $ ^2F_{7/2}  $& 737    & -468 &  -353    &    -363    &    -319    &    -380    \\
       & $ 5f^26p $& $ ^2F_{5/2}  $& 464    & -730 &  -722    &    -729    &    -766    &    -748    \\
       & $ 5f^26p $& $ ^2G_{7/2}  $& 512    & -584 &  -655    &    -662    &    -683    &    -681    \\
       & $ 5f^26p $& $ ^4I_{11/2} $& 425    & -781 &  -762    &    -768    &    -814    &    -787
   \end{tabular}
\end{ruledtabular}
\end{table}
%%%%%%%%%%%%%%%%%%%%%%%%%%%%%%%%%%%%%%%%%%%%%%%%%%%%%%%%%%%%%%%%%%%%

\begin{table*}[tb]
\caption{\label{qed2}
%%%%%%%%%%%%%%%%%%%%%%%%%%%%%%%%%%%%%%%%%%%%%%%%%%%%%%%%%%%%%%%%%%%%
Transition energies (cm$^{-1}$) for Ba$^{8+}$\!, Eu$^{14+}$\!, and
Cf$^{15+}$ calculated using the CI+all-order method and M1 version
of QED potential. Experimental results for Ba$^{8+}$ are from Ref.\
\cite{NIST}. Columns QED, 3e, and Total present QED corrections,
contribution of the effective three-electron interactions
\cite{KSPT16}, and final theoretical values, respectively.
}
\begin{ruledtabular}
\begin{tabular}{llrrrrlllrrrrlllrrr}
 \multicolumn{1}{c}{Conf.}& \multicolumn{1}{c}{Term}&
 \multicolumn{1}{c}{Expt.}& \multicolumn{1}{r}{QED}&
 \multicolumn{1}{r}{Total}& \multicolumn{1}{r}{Diff.}
 & &
 \multicolumn{1}{c}{Conf.}&
 \multicolumn{1}{c}{Term}& \multicolumn{1}{r}{CI+all}&
 \multicolumn{1}{r}{QED}& \multicolumn{1}{r}{3e}&
 \multicolumn{1}{r}{Total}
 & &
 \multicolumn{1}{c}{Conf.}&
 \multicolumn{1}{c}{Term}& \multicolumn{1}{r}{QED}&
  \multicolumn{1}{r}{3e}&\multicolumn{1}{r}{Total}\\
\hline
\\[-3mm]
\multicolumn{7}{c}{Ba$^{8+}$}& \multicolumn{6}{c}{Eu$^{14+}$}&
\multicolumn{6}{c} {Cf$^{15+}$}\\  \hline
\\[-3mm]
%\hline
$5s^2$&$^1S_0$&      0 &$   0$&     0  &$       $&&$4f^25s$& 3.5 &     0 &$   0 $&$  0 $&     0 &&$5f6p^2$&$ ^2F_{ 5/2} $ &$   0$&$  0$&      0  \\
$5s5p$&$^3P_0$& 116992 &$-496$& 117769 &$0.66\% $&&$4f^3  $& 4.5 &  3161 &$-1199$&$-700$&  1262 &&$5f^26p$&$ ^2F_{ 9/2} $ &$-524$&$119$&  12898  \\
$5s5p$&$^3P_1$& 122812 &$-491$& 123492 &$0.55\% $&&$4f^25s$& 4.5 &  2594 &$-1   $&$  1 $&  2594 &&$5f6p^2$&$ ^2F_{ 7/2} $ &$-115$&$-18$&  22018  \\
$5s5p$&$^3P_2$& 142812 &$-455$& 143661 &$0.59\% $&&$4f^3  $& 5.5 &  7275 &$-1198$&$-689$&  5388 &&$5f^26p$&$ ^2F_{ 5/2} $ &$-484$&$ 29$&  27127  \\
$5s5p$&$^1P_1$& 175712 &$-457$& 175683 &$-0.02\%$&&$4f^25s$& 5.5 &  6699 &$ 1   $&$ -4 $&  6696 &&$5f^26p$&$ ^2G_{ 7/2} $ &$-416$&$-45$&  29214  \\
$4f5s$&$^3F_2$& 237170 &$-530$& 236939 &$-0.10\%$&&$4f^25s$& 1.5 &  9705 &$ 1   $&$ -3 $&  9703 &&$5f^26p$&$ ^4I_{ 11/2}$ &$-523$&$ 48$&  37081  \\
$4f5s$&$^3F_3$& 237691 &$-530$& 237457 &$-0.10\%$&&$4f^3  $& 6.5 & 11513 &$-1197$&$-683$&  9633 &&$5f^26p$&$ ^4H_{ 9/2} $ &$-528$&$ 37$&  37901  \\
$4f5s$&$^3F_4$& 238547 &$-530$& 238294 &$-0.11\%$&&$4f^25s$& 2.5 & 11300 &$ 1   $&$ -3 $& 11298 &&$5f^26p$&$ ^4G_{ 7/2} $ &$-511$&$ 54$&  40206  \\
$4f5s$&$^1F_3$& 245192 &$-544$& 245280 &$0.04\% $&&$4f^25s$& 6.5 & 11420 &$ 3   $&$ -9 $& 11414 &&$5f^26p$&$ ^2D_{ 5/2} $ &$-525$&$ 45$&  42287%\\[-0mm]
\end{tabular}
\end{ruledtabular}
\end{table*}

%%%%%%%%%%%%%%%%%%%%%%%%%%%%%%%%%%%%%%%%%%%%%%%%%%%%%%%%%%%%%%%%%%
In Tables~\ref{alkali} and ~\ref{Cu-like} we compare the SE values
obtained using methods M1, M2, M3, and M4 described above with the
\textit{ab initio} calculations of Refs. \cite{sap_02} and
\cite{che_06} respectively, to which we refer as ``exact''.

Calculations of the SE shifts in Refs. \cite{sap_02,che_06} were
performed with the local potential $V_{\rm eff}(r)$:
\begin{eqnarray}
V_{\rm eff}(r)=V_{\rm nuc}(r) - \!\int \limits_0^{\infty}\!
dr'\frac{\rho(r^{\prime})}{r_{>}}
+x_{\alpha}\Bigl[\frac{81}{32\pi^2}r\rho_{\rm}(r)\Bigr]^{1/3}\!\!,
\end{eqnarray}
where $V_{\rm nuc}(r)$ is nuclear potential and $\rho_{\rm}(r)$ is
total electron charge density. The choice $x_{\alpha}=0$ corresponds
to the Dirac-Hartree potential, $x_{\alpha}= 2/3$ is the Kohn-Sham
potential, and $x_{\alpha}= 1$ is the DFS potential.
%The details of calculation of the local potential $V_{\rm eff}$ are
%described in \cite{sap_02, che_06}.
Our data were obtained by averaging the SE operator $V^{\rm SE}$ with
the wave function of the valence state determined from the Dirac equation
with the potential $V_{\rm eff}(r)$.

In Table \ref{alkali}, the SE shifts for the valence s-state of the
neutral alkali atoms are given in terms of function $F(\alpha Z)$,
defined by
\begin{eqnarray}
\Delta E^{\rm SE}=\frac{\alpha}{\pi}\frac{(\alpha Z)^4}{n^3}F(\alpha Z) \,
mc^2 \,.
\end{eqnarray}
Table \ref{alkali} illustrates that the  SE shifts obtained using
M1, M2, and M4 methods are in very good agreement with exact
results. We find some discrepancies between the data calculated
using the local radiative potential (method M3) and exact values,
especially for low Z.

In Table \ref{Cu-like} we present the SE corrections calculated for
the $4s-4p$ and $4p-4d$ transition energies of Cu-like ions. The
results obtained within methods M1, M2, and M4 are in very good
agreement with the exact ones. There is slight deviation of the data
obtained in method M3 for high Z. Note that method M3 was recently
modified in Ref. \cite{Ginges_16b}, where more complicated and
accurate finite size correction to the radiative potential and
additional fitting for $d$ states were introduced.

Comparison of the QED corrections to the energies of Ba$^{8+}$,
Eu$^{14+}$, and Cf$^{15+}$  obtained using all four QED potentials
is given in Table~\ref{qed1}. The results in column labeled
CI-M1 are obtained by including the QED potential only in the CI
Hamiltonian, using the first variant of the QED potential. In this
version of the calculations, the finite basis set is constructed
with no QED corrections. Respectively, the QED corrections for the
$4f$ and $5f$ orbitals are zero owing to no overlap with the
nucleus. In all other calculations QED potential is added in both
CI~Hamiltonian and in the construction of the basis set, which
effectively changes the $nf$ orbitals via the modification of the
self-consistent potential.
The results of such calculation are listed in column labelled
 CI-M1. Comparison of  these values with full QED calculations (column M1)
   shows that while the differences between these approaches
are minor for Ba$^{8+}$, they are very significant for heavier ions
with higher degree of ionization. When the QED contribution to the
ground state is subtracted, the differences between CI-M1 and M1
approaches are still significant, 5\%, 14\%, and 25\%  for Ba$^{8+}$,
Eu$^{14+}$, and Cf$^{15+}$, respectively.

We also carried out the same calculations using the less accurate method
that combines CI and
%many-body perturbation theory (MBPT)
MBPT \cite{KozPorSaf15} to evaluate if the accurate treatment of the
electronic correlation is important for the QED calculation. In the
CI+MBPT method, core-valence correlation are treated in the second
order of MBPT.
CI+MBPT results are listed in column labelled M1$^{\prime}$
The differences between the QED contributions calculated in the
CI+MBPT and CI-all-order methods are small for Ba$^{8+}$ and
Eu$^{14+}$, but significant for $J=\tfrac72$ $5f6p^2$ and $5f^26p$
levels of Cf$^{15+}$. These $J=\tfrac72$ levels are strongly mixed
and all-order corrections change weights of $6p$ and $5f$ electrons
in the many-electron wavefunction, which affects the QED
contributions.
The differences between the calculations carried out with different
QED potentials are generally small, with the biggest difference for
the QED~M3 potential. The differences increase for Cf$^{15+}$, where
QED corrections are the largest.

The QED corrections to the energies  of Ba$^{8+}$, Eu$^{14+}$,
Cf$^{15+}$ calculated using the CI+all-order method with the first
version of the QED potential are given in Table~\ref{qed2} to show
the relative size of the QED corrections to the energy levels.
All values are given relative to the corresponding ground state.
%are subtracted from the corresponding ground-state configuration.
Final values that include QED corrections are given in columns
``Total''. Non-QED part of the calculation is the same as in
%values are the same as in
\cite{SafDzuFla14,SafDzuFla14a,SafDzuFla14b,DzuSafSaf15}. Our final
results for Ba$^{8+}$ are in excellent agreement with experiment
\cite{NIST}. The QED corrections are very significant for low-lying
$4f^3$ levels of Eu$^{14+}$, so we have also included the
CI+all-order values without QED for clarity.

%3e correction>>>
In the CI+MBPT and CI+all order calculations for the systems with
three or more valence electrons there is an additional contribution
to the valence energy from an effective three-electron (3e) interaction
between valence electrons \cite{DFK96}. This contribution may be
enhanced for the systems with an open $f$ shell \cite{KSPT16}.
Respective 3e corrections appear for Eu$^{14+}$ and Cf$^{15+}$ ions and
are listed in Table~\ref{qed2}.
These corrections are comparable to QED corrections for Eu$^{14+}$.
%3e correction<<<

In summary, we find that accurate
treatment of the QED effects is essential for reliable prediction of
the transition energies in HCIs with optical transitions of interest
to the clock development and tests of fundamental physics.
The QED corrections in these ions  are large enough to significantly affect the predictions
of the transition wavelengths.
Our results show that the QED
corrections obtained by all four QED potentials
 are very similar, with the difference being smaller than
the estimated uncertainty in the treatment of the correlation
correction. We find that it is imperative to include the QED
correction both in the construction of the basis set orbitals and
into the CI Hamiltonian, in particular for the configurations
involving $5f$ electrons, as in the example of Cf ions. In the case
of strong configuration mixing, QED corrections calculated in the
CI+MBPT (M1$^\prime$) and CI+all-order (M1) approximations may
differ by as much as 100~cm$^{-1}$.  We
demonstrate that QED effects can be reliably accounted for by
incorporating the modern  QED potentials
into the CI+all order method. Finally, high precision calculations
of the systems with more than two valence electrons should include
contribution of the effective three-electron interactions between
valence electrons together with QED effects.

This work is partly supported by the Russian Foundation for Basic
Research Grants No. 14-02-00241, 15-03-07644, and No.16-02-00334, by
U.S. NSF grant No. \ PHY-1520993 and SBbSU Grants No:
11.38.269.2014, 11.38.237.2015, and 11.38.261.2014.

%\clearpage
%%%%%%%%%%%%%%%%%%%%%%%%%%%%%%%%%%%%%%%%%%%%%%%%%%%%%%%%%%%%%%%%%%%%%%%%%

\end{document}